# Computational Group Selection

Nripsuta Saxena

April 2021

**Abstract**

Humans spend a significant part of their lives being a part of groups. In this document we propose research directions that would make it possible to computationally form productive groups. We bring to light several issues that need to be addressed in the pursuit of this goal (not amplifying existing biases and inequality, for example), as well as multiple avenues to study that would help achieve us this task efficiently.

## 1  Introduction

Human beings are social animals. In almost every phase of life we interact with other humans. We study and party with our friends, work in teams at work, and form groups for leisure activities. A significant part of our lives is spent being associated with groups. Sports teams, teams at work, groups for class projects, and so forth–it is no coincidence that so much of what we do is achieved by being part of a group. Research has consistently shown that groups of people working together can be more effective and productive than individuals working on their own [3]. Research has also shown that not every group of people will work well together [7]. There are certain characteristics good teams or productive groups have that make them better than other groupings of people.

All groups are currently formed by people. Imagine the number of hours each person spends over the course of their lives trying to figure out which groups or teams to be a part of! A lot of people have to spend their valuable work-hours on conundrums like this too. Organizations (including academic institutions and companies) have to figure out which people to make offers to – who will work well with each other, be productive and efficient?

[3]

What if we could computationally design efficient, productive teams that are bigger-than-the-sum-of-its-parts? Admissions committees will not have to devote hundreds of work-hours each year on trying to achieve the ultimate combination of students in its incoming class. Program and product managers in companies don't have to spend any time trying to gauge which candidates might gel well with their existing team. Professors can focus on their research, and managers on their teams and projects. This is in addition to the very obvious benefit of understanding exactly what it is that makes a group of people bigger than the sum of their parts.

This is our goal. To design a model that can select a group of people who possess the desired characteristics and skills, and are together bigger than the sum of their parts. This can be the scenario where you get to choose the entire team (eg. a school's incoming class), or selecting a few people to add to an existing team (eg. hiring a software developer for an existing product team).

## 2 Background

Then there is also the issue of bias seeping into the process. Most people have implicit biases. Research has shown that, at least in the field of hiring, people making the hiring decisions tend to prefer candidates with a background similar to their own. [11] found that employers preferred hiring candidates who were not only competent but also similar to themselves in preferences for leisure activities and experiences. These cultural preferences often *outweighed* concerns about productivity. It is not hard to see how why this might be a problem. The author observed hiring decisions being made at professional service firms (law, investment, consulting firms), and found that given two equally qualified candidates, employers tend to prefer the applicant who is similar to them in terms of how they choose to spend their time *outside* of work. Unsurprisingly, since most employees at such firms tend to be from privileged backgrounds, their hobbies tend to be expensive. If the people making the hiring decisions care about whether you like to golf or play the cello like they do, applicants whose families could not afford such expensive pursuits growing up will be at a disadvantage.

[12] found that social class also matters in hiring in elite professions. Applicants who were perceived to be higher-class and male received significantly more callbacks than higher-class women, and lower-class men and women. Higherclass men are seen as a 'better fit' with the elite culture and clientele of large professional firms. Higher-class women do not see this boost because of the negative stereotype that they will not be as committed to a full-time career.

Research [10] has also shown that elite employers screen applicants based on the selectivity of the educational institution the applicants attended. The more selective the university, the better. Being from an extremely prestigious, selective university was not enough on its own. After screening for university, elite employers then screen for for extracurricular accomplishments. Unsurprisingly, the most favorable activities tend to be extremely expensive to maintain an interest in, making it extremely likely that the applicant is from a well-off socioeconomic background.

Race can also significantly affect an individual's employment prospects. Researchers designed an experiment where [9] equivalent resumes for white, black, and Latino applicants matched by demographics and interpersonal skills were sent to hundreds of entry-level jobs. Equally qualified black applicants were *half as likely* as their white counterparts to receive a callback or a job offer.

Black and Latino applicants with no criminal records did as well as white applicants *with criminal records!* Other studies have also found that white applicants *with* a criminal record received more callbacks and better treatment than black applicants *without* a criminal record [8].

Even in university admissions, applicants from high-income families have a significant advantage. They are able to get tutoring for standardized tests like the SAT, and are able to retake these tests multiple times, sending universities only their best score [2]. Applicants from low-income families typically take such tests only once, and usually do not have access to any tutoring.

So applicants from higher-income, higher-status families have a significant advantage while applying to college, and this advantage is further reinforced when they go through the recruitment process for jobs. Things are even more complicated for people of color.

## 3 Where We Are

In today's world this task is solved almost entirely manually. University admissions (or job) applications tend to be filtered computationally first according to some criteria (eg. GPA, number of years of experience, etc). After that, they tend to go through multiple rounds of review by the admissions (or hiring) committee. The number of eligible applicants is reduced in each round of review. Interviews might also be conducted before final decisions are made.

When you consider how most competitive schools get tens of thousands of applications each year (Harvard received more than 57,000 applications in 2021, and over 40,000 applications each year since 2018 [5]), and how competitive jobs might get hundreds of job applicants per posting, it becomes clear that the amount of human work-hours spent on these reviewing processes is enormous.

This is obviously not an ideal situation.

There hasn't been much work so far in this research area. [4] explore how collaboration can lead to a team of individuals utilizing their skills as effectively as possible, and builds on literature from the fields of sports and organizational psychology. They focus on which individuals might work well together (which they refer to as team collaboration structures), and interaction modes (i.e. how people choose to interact with one another). They however choose not to explore the question of team size, and limit their focus to teams of five persons with specific collaboration structures. They formulate collaboration structures such that within the team of 5 people, groups of two individuals may choose to interact with each other and not with others (there may be 11 such dyads), or groups of three may choose to work together and exclude the remaining two. The work also assumes that each individual might choose to rely only on their own cognitive ability to get their work done, or they may choose to work collaboratively with others. So while they may be working towards a common goal, they may or may not be relying on someone else and work inter-dependently (rather than independently) to help them accomplish the goal. The present work provides a nuanced understanding of how team collaboration leads to team members utilizing their most effective skills to impact team performance.

[6] look at how to select the optimal size and interaction frequency for a team. The authors also explore how interacting teams (teams which work interdependently and collaboratively to achieve the common goal) performed compared with teams where members worked independently to complete their contributions to the bigger goal.

Outside of these two works, we are not aware of work that tries to take a computational approach to team formation.

## 3.1 Limitations

### 3.1.1 Subtle difference between University sceraio and workplace scenario

We feel this issue needs to be explored in more depth, and from different perspectives. For example, so far previous work has only focused on computationally studying team formation in a workplace setting. In other words, they look at the problem of what makes a team in a workplace effective in achieving a certain task, like a deadline for a product deliverable? But the goal of studying groups in a university is not the same. In a university setting, the aim would be to put together a group of people who will be able to learn from the experiences of each other [1]. For example, future lawyers will benefit from studying in a diverse environment and being able to think about facts and arguments from many angles [1].

### 3.1.2 Constraints

Work exploring this problem so far has done so in very specific constraints. Team size has been extremely limited, as well as the assumptions about how the team members interact with each other. We believe for work in this area to be practically useful there need to be fewer constraints, at least in terms of team size.

We might need to add different types of constraints as well. For example, if a company needs to hire a team to develop a product they might want at least two software developers, one product manager, one sales person, etc. Similarly, a university might have constraints they need to satisfy when selecting their incoming class. For example, they may need a certain number of athletes for replacing current student athletes who are about to graduate, some debaters for the debate team, they may like having students from different parts of the country and the world to create diverse experiences on campus, etc.

---

[1] Sweatt v. Painter, 339 U.S. 629, 634 (1950) "[L]legal learning [is ineffective in isolation from the individuals and institutions with which the law interacts.).

# 4     Questions at the Research Frontier

We recommend a sizeable chunk of time be first devoted to learning the existing literature on team-building, admissions, and hiring. Data should then be the focus of attention. Only after researchers have a solid understanding of the background and what data hiring/admissions officers typically pick up on should the researchers move on to building a system and testing. It may seem as though there is nothing really happening for quite a while, but having a solid foundation of the knowledge collected so far is extremely important.

## 4.1     What makes a group bigger-than-the-sum-of-its-parts

As past research has shown, not every grouping of people is productive. So what traits or characteristics make a en efficient group click? A lot of research exploring this question has focused on dyads (i.e. teams of two people), and not much academic research has explored what makes an incoming class biggerthan-the-sum-of-its-parts. A good starting point would be to conduct studies including interviews with admissions committee members and seasoned hiring managers, as well as observing the entire process as a quiet onlooker.

Once we have a clear understanding of what characteristics, traits, or other pointers seasoned professionals look at, it will help us figure out the variables that matter most in this problem.

## 4.2     What data will we need for this?

The next step would be to figure out what data needs to be collected. When a hiring manager or an admissions committee member decides which applicant would be a good fit for a team or an incoming class, what things are they looking for? What factors help them judge whether one applicant will be a better fit over another? How do admissions officers decide which students to accept, and which ones to place on a waiting list? Researchers working on this problem will first have to recognize what facts and traits these experienced hiring/admissions committee members are picking up on, and make sure they somehow include that in the data they collect. This would also involve exploring how often the decision-makers' intuition about who would be a good fit actually turns out to be correct. Admissions and hiring managers don't get to observe the counterfactual, so they have no way of learning whether the applicants they rejected might have turned out to be a good fit after all. How often on average do hiring managers regret hiring a certain applicant? What do they learn from those experiences, and how do they refine their hiring process? While it might be easy to evaluate whether a certain employee is doing a good job and gets along well with the rest of their team, evaluating students in an incoming class may be much harder. How do universities decide whether a class has indeed been greater than the sum-of-its-parts?

Then there's the question of what an interview manages to capture. Apart from testing the skills of an applicant, interviews are often used to judge the hard-to-define "fit." Of course, care should be taken to ensure that doesn't always translate to cultural matching [11].

### 4.3 Human-in-the-loop direction

A direction worth exploring is the human-in-the-loop model. It is likely that it would be beneficial to include some form of human feedback in the model, perhaps via having admissions/hiring managers interview the candidates the model thinks are the most appropriate. This might be a worthwhile approach to incorporating some of the intuition some admissions/hiring managers say they rely on while making admissions/hiring decisions. Of course, extreme care would have to be given to ensure the biases mentioned above don't creep in as a result of this.

### 4.4 Equity constraints

This is perhaps the most important question at the research frontier. There is research going back decades and decades that shows these domains can be heavily biased, and we want to be sure there is as little of these biases as possible in working towards building an efficient computational group-selection model. It is also worth pointing out that we can only look (and try to correct) for biases that we are currently *aware* of. Involving as many varied stakeholders as possible in the development and testing process will likely be helpful.

### 4.5 Not contributing to the problem ourselves

Care must also be given to ensure the system itself doesn't introduce any unintended consequences. Such models will have consequences for people, in some cases long-term, life-changing consequences. This is a massive responsibility, and researchers would do well to be cognizant of this at all stages of the process.

It might be a good idea to have fair, strict stakeholders overseeing the process to ensure we don't unintentionally amplify existing systemic inequality and privilege.

Societies have been trying to reduce inequality, inequity, and injustice for centuries and we still live in an extremely imperfect, unfair society so this is not going to be an easy problem to address. In all probability everything we do to address this will be insufficient. But it is *extremely important* we build in mechanisms that enable us to make appropriate changes as and when we become aware of problems.

### 4.6 How do we test it?

Once researchers have managed to create the dataset, and we can use it to optimize our equations, there will need to be work on figuring out how to test it. How do we judge if a team at work is productive? If a graduating class has achieved what we expected them to achieve? What metrics do we use?

Some standard metrics for a class might be average graduation rate, how many students had jobs or offers for grad school lined up upon graduation. For a team in the workplace, targets might be directly related to the job description (eg. $x amount in sales, new product launch, etc). It might be easier to test a model in a workplace setting, since in a university setting researchers might need at least about four years to be able to measure the metrics (graduation rates, etc).

## 4.7 How do we incorporate feedback/effects from the real world?

Another consideration to look at is that this should be framed as a one-shot problem. The effects this model will have on people and society in general will shape the applicant pool of the future. Researchers would do well not to ignore this fact and try to incorporate this into refining their model over time.

This is not going to be an easy thing to figure out since currently most equity- and fairness-related research has also framed problems as one-shot scenarios. But until we learn how to address this very real phenomenon, substantial progress in this domain will be hard to make. Until we understand more about how to do this, it will also be quite hard to get such systems adopted in the real world.

## 4.8 Collaboration: Taking advantage of extant literature

The best way to approach such a problem might be as a collaborative venture between computer scientists and social scientists, as well as economists. Social scientists and economists have been studying these issues for decades, and not learning from them and their experience would be a massive lost opportunity.

# 5 Conclusion

In this paper we aim to highlight an understudied problem that might help improve the formation of good teams while simultaneously reducing the number of human work-hours that are currently spent on making these decisions. We discuss some glaring issues that need to be addressed and point to some interesting research directions for future work while also highlighting the need for vigilant and strict oversight of this process so as not to unintentionally amplify current biases or introduce new ones.